\documentclass[twocolumn,showpacs,superscriptaddress,preprintnumbers,amsmath,amssymb,epsfig,floatfix,prl]{revtex4-1}
\usepackage{epsfig}
\usepackage{graphicx}
\usepackage{dcolumn}
\usepackage{color}
\usepackage{natbib}  %%%%%
\definecolor{blue}{RGB}{112,196,255}
\usepackage{bm}
\usepackage{tabularx}
\newcolumntype{L}[1]{>{\raggedright\arraybackslash}p{#1}}
\newcolumntype{C}[1]{>{\centering\arraybackslash}p{#1}}
\newcolumntype{R}[1]{>{\raggedleft\arraybackslash}p{#1}}
\begin{document}
\title{Kinetics of Topological Stone-Wales Defect Formation in Single Walled Carbon Nanotubes}	
\author{Mukul Kabir}
\email{mukul.kabir@iiserpune.ac.in} %%%%%
\affiliation{Department of Physics, and Centre for Energy Science, Indian Institute of Science Education and Research, Pune 411008, India}
\author{Krystyn J. Van Vliet}
\affiliation{Department of Materials Science and Engineering, Massachusetts Institute of Technology, Cambridge, Massachusetts 02139, USA}
\date{\today}
%\keywords{carbon nanotube, topological defect, formation and activation energy, doping}

%\begin{document}
\begin{abstract} 
Topological Stone-Wales defect in carbon nanotubes plays a central role in plastic deformation, chemical functionalization, and superstructure formation. Here, we systematically investigate the formation kinetics of such defects within density functional approach coupled with the transition state theory. We find that both the formation and activation energies depend critically on the nanotube chairality, diameter, and defect orientation. The microscopic origin of the observed dependence is explained with curvature induced rehybridization in nanotube. Surprisingly, the kinetic barrier follows an empirical Br{\o}nsted-Evans-Polanyi  type correlation with the corresponding formation energy, and can be understood in terms of overlap between energy-coordinate parabolas representing the structures with and without the defect. Further, we propose a possible route to substantially decrease the kinetic activation barrier. Such accelerated rates of defect formation are desirable in many novel electronic, mechanical and chemical applications, and also facilitate the formation of three-dimensional nanotube superstructures. 

\end{abstract}
\maketitle
\section{Introduction}
% Why study defects in carbon nanostructures? 
Defects including atomic impurities, vacancies, or topological junctions and kinks may be present in as-prepared carbon nanotubes (CNTs)~\citep{doi:10.1021/ar010166k, CNTDefectChapter} and graphene.\citep{doi:10.1021/nn102598m} 
%What kind of defects are we interested in?
In particular, Stone-Wales (SW) defects are important topological defect class,~\citep{Stone1986501, Wales.Nature.394} with analogy to dislocation dipoles in bulk materials. 
Individual SW defects have been identified experimentally in fullerene,~\citep{Nat.Mater.7.790} CNTs,~\citep{NatureNanotech.2.358,PhysRevB.83.245420,doi:10.1021/ja100760m} and graphene.~\citep{doi:10.1021/nn102598m,doi:10.1021/nl801386m}

%Why SW defects are important to study?
The presence of SW defects markedly alters the chemical and mechanical properties of CNTs.  Moreover, abundance of these SW defects is desirable for many novel chemical, and electronic applications requiring CNT network formation.~\cite{doi:10.1021/nl015541g, PhysRevB.66.245403, PhysRevLett.92.075504, NatMatKrasheninnikov, doi:10.1021/cr050569o, doi:10.1021/nn700143q,ANIE:ANIE200705053,Scientific.Report.2}  For example, the SW transformation is found to be the microscopic unit process for nanojunction formation.~\citep{doi:10.1021/nl015541g, PhysRevB.66.245403, PhysRevLett.92.075504,  NatMatKrasheninnikov} Additionally, the chemical reactivity and electronic properties of CNTs~\citep{doi:10.1021/cr050569o} and graphene~\citep{doi:10.1021/nl802234n} are modified by presence of SW defects, which act as anchor sites for chemical functionalization and are thus desirable in that context.  It has also been posited that plastic deformation of CNTs is mediated via spontaneous formation and migration of topological SW defects under external tension.~\citep{NatureNanotech.2.358,PhysRevLett.81.4656, PhysRevB.57.R4277} Thus, the thermodynamics and kinetics of the SW defect pose multiple implications for understanding and use of CNTs. 

These properties of the SW defect are already established for graphene. The thermodynamic formation energy and the concurrent kinetic barrier for SW formation are very high at $\sim$ 5 and 10 eV, respectively, in graphene.~\citep{doi:10.1021/nn102598m, PhysRevB.80.033407} However once formed, the high reverse kinetic barrier ($\sim$5 eV) implies defect stability over a wide temperature range. In contrast to graphene, it could be anticipated that both the formation energy and kinetic barrier for CNTs will depend explicitly on intrinsic structural parameters including the CNT diameter, and the relative orientation of SW defects. However, these dependences have not yet been determined. In this Letter, we analyze both the thermodynamics and kinetics of SW formation in single-wall CNTs.  We address systematically how these quantities depend on nanotube chirality and diameter, as well as relative defect orientation. We also correlate the thermodynamic formation energy to the activation barrier, and provide a microscopic description.  Further, we show that the SW activation barrier can be modified substantially by substitutional heteroatom doping. Such doping accelerates the SW formation by six to twenty orders of magnitude at relevant temperatures and, in turn, assists nanotube welding~\citep{doi:10.1021/nl015541g, PhysRevB.66.245403, PhysRevLett.92.075504, NatMatKrasheninnikov} and formation of three-dimensional nanotube superstructures as has been observed experimentally.~\citep{doi:10.1021/nn700143q,ANIE:ANIE200705053,Scientific.Report.2}

\begin{figure*}[t]
\centering
\includegraphics[scale=0.07]{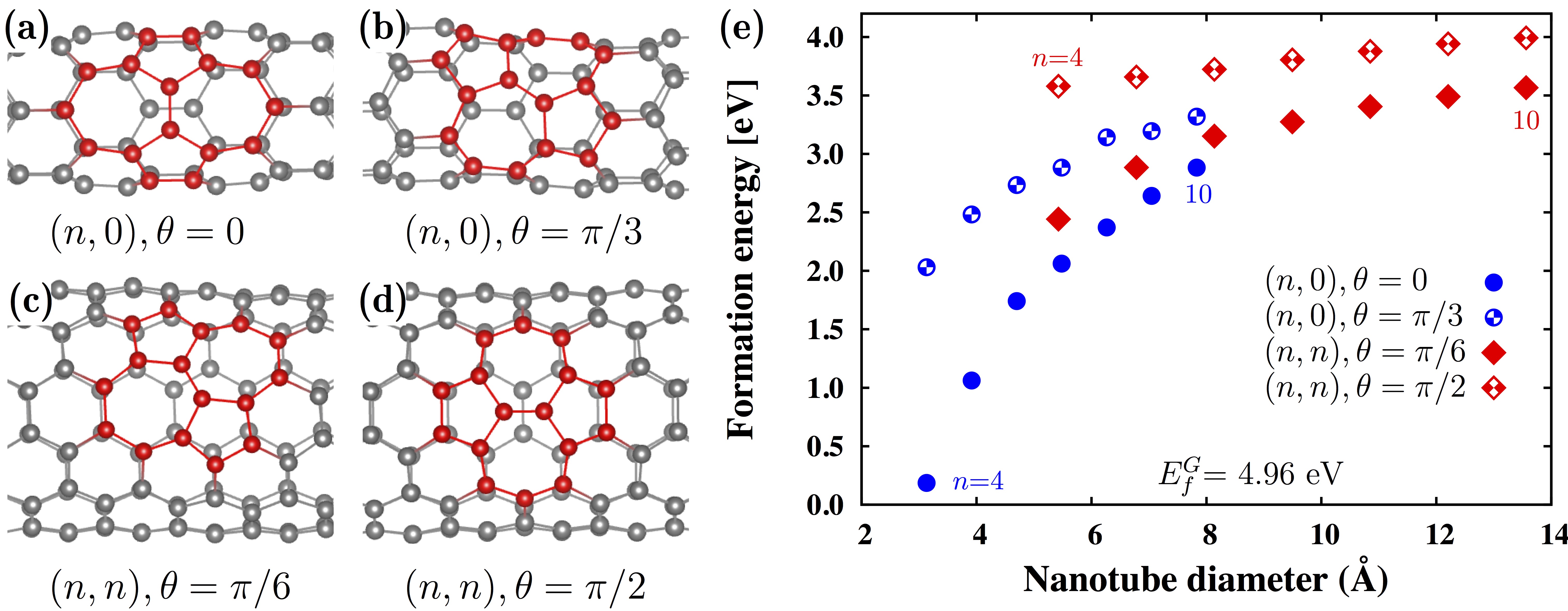}
\caption{Orientation of SW defects strongly influences defect formation and activation energies. Among three possible orientations, two are inequivalent and are shown for optimized $(n,0)$ zigzag [(a) and (b)], and $(n,n)$ armchair [(c) and (d)] nanotubes. Here, $\theta$ is defined as the angle between the rotating C--C bond in the pristine CNT with the tube axis. The CNT diameter is related to the chiral vectors $(n,m)$: $d = (\sqrt{3}a/\pi)\sqrt{n^2+m^2+nm}$, where $a$ is the C--C bond length (1.425 \AA). The variation of SW formation energy is shown in (e), which depends crucially on the type of CNT and SW orientation $\theta$; this energy increases with nanotube diameter $d$ and approaches the magnitude for two-dimensional graphene (4.96 eV). This trend substantiates the effect of the local curvature at the rotating C--C bond on the defect formation energy, which can be explained via curvature induced rehybridization in CNTs. Calculated zero point energy correction $\Delta$ZPE to the formation energy is always $<$ 0.1 eV.}
\label{fig:SWf}
\end{figure*}

\section{Computational Details}
Here we couple density functional theory (DFT) and the climbing-image nudged elastic band (CINEB) method that correctly predicts the first-order transition state.~\citep{PhysRevB.47.558,PhysRevB.54.11169,Henkelman-1}  The DFT calculations are carried out within the projector augmented wave potential~\cite{PhysRevB.50.17953} implemented in the Vienna {\em ab initio} simulation package~\citep{PhysRevB.47.558,PhysRevB.54.11169}. A plane-wave cutoff of 500 eV, and Perdew-Burke-Ernzerhof generalized gradient approximation are used.~\citep{PhysRevLett.77.3865} The first Brillouin zone is sampled with a Monkhorst-Pack grid~\citep{PhysRevB.13.5188} of 1$\times$1$\times$4. The position of all atoms are relaxed until the forces are less than 0.01 eV/\AA. The tube axis is along the $Z$-direction, and a vacuum space of more than 10 \AA \ along the $X$ and $Y$ directions is used to eliminate the image interaction. Minimum energy path for SW defect formation were sampled using the CINEB method.~\citep{Henkelman-1} In CINEB, a set of intermediate structures (images) are distributed along the reaction path connecting optimized pristine and defected carbon nanotubes (CNTs). The images are connected via an elastic spring, and each intermediate image is fully relaxed in the hyperspace perpendicular to the reaction coordinate. The nature of the transition states have been confirmed via the phonon calculation, where one and only one imaginary frequency confirms the transition state to be of first-order. 

The SW defect produces a long-range strain field, and the resultant dislocation-dislocation interaction may affect formation and activation energies. We indeed find that these energies depend on the CNT length, and converge at tube length of $\sim$2.5 nm, which we fixed throughout the present calculations (Table S1 in Supporting Information).  On a CNT surface, the SW defect can have three possible orientations: $\theta =$ $\chi$ and $\pi/3 \pm \chi$, where $\chi$ is the chiral angle.~\citep{zhou:1222} Of these three sets, two inequivalent orientations are shown in Fig.~\ref{fig:SWf}(a)-(b) for zigzag, and Fig.~\ref{fig:SWf}(c)-(d) for armchair nanotubes. The formation energy is calculated as $E_f=E_{\rm SW} - E_{\rm P}$, where $E_{\rm SW}$ and $E_{\rm P}$ are the energies of the CNT with and without the defect, respectively. The activation barrier is calculated as the energy difference between the first-order transition state and the pristine CNT.  Although few attempts have been made to calculate the formation energy alone,~\citep{zhou:1222, dumitrica:2775} there is no systematic investigation that facilitates comparisons or general conclusions. Moreover, prior attempts either suffered from inadequate chemical accuracy for CÐC bonding (classical many-body potential or tight-binding approach)~\citep{PhysRevLett.81.4656, PhysRevB.57.R4277, zhou:1222} or assumed insufficient structural models,~\citep{dumitrica:2775} and did not consider the plausible orientational contribution.

\section{Results and Discussion}
{\bf Formation energy, nanotube curvature, and defect orientation.} The calculated SW defect formation energy depends strongly on the nanotube diameter $d$, and the orientation of SW dislocation dipole $\theta$.  Figure~\ref{fig:SWf}(e) illustrates a systematic variation of $E_f(d, \theta)$ obtained with our present calculations, from which we observe two distinct trends. 
%Observation - I
First, the calculated $E_f(d, \theta)$ increases monotonically with increasing $d$ for any particular $\theta$, and converges toward the value for two-dimensional graphene $E_f^G$. Applying the identical theoretical approach for graphene, $E_f^G$ is calculated to be 4.96 eV, which agrees well with previous calculations.~\citep{PhysRevB.80.033407} Note that the SW defect formation energy (1.57 eV) for C$_{60}$ fullerene (diameter $\sim$ 7 \AA) is comparable with that of a nanotube with similar diameter.~\cite{PhysRevB.84.205404} 
%Observation - II
Interestingly, we find that for any particular CNT with diameter $d$, the calculated $E_f(d, \theta)$ increases monotonically with the angle $\theta$ made by the rotating C--C bond with the tube axis in the pristine structure [Fig.~\ref{fig:SWf}(e)]. For zigzag and armchair nanotubes, we find the formation energy to follow $E_f[(n,0), \theta=0] < E_f[(n,0), \theta=\pi/3]$ and  $E_f[(n,n), \theta=\pi/6] < E_f[(n,n), \theta=\pi/2]$ order. 
%Explanation - General intro
Since $E_f(d, \theta)$ is the energy of the defected structure relative to the corresponding nanotube without the defect, the dependence on $d$ and $\theta$ can be explained qualitatively by the curvature induced rehybridization for the defected structure (see Supporting Information).~\cite{Haddon17091993, Dumitrica2002182} The Coulomb repulsion inside the nanotube increases with increasing curvature, leading to significant rehybridization between $\pi$ and $\sigma$ orbitals. Thus, the true hybridization in CNTs is intermediate between $sp^2$ and $sp^3$, i.e., $sp^{2+\tau}$ with $\tau \in$ [0,1] is the degree of rehybridization. 
%Explanation specific to diameter dependance 
With increasing diameter (decreasing curvature), $\tau$  decreases rapidly and approaches zero, and the hybridization state of the affected bond is increasingly $sp^2$-like (Table S2 in Supporting Information). Thus, relative to the pristine structure, the energy of the defected CNT shifts toward higher energy with increasing $d$ (Fig. S3 in Supporting Information). Therefore, $E_f(d, \theta)$ increases with increasing $d$, and approaches to $E_f^G$ of two-dimensional $sp^2$-graphene. 
%Explanation specific to angle dependance
Similarly, for a given $d$, the $\theta$ dependence can be explained by considering the local environment of the rotated C--C bond for the defected structure. With increasing $\theta$, the local curvature of the rotated C--C bond [shown in Fig.~\ref{fig:SWf}(a)-(d)] decreases, and thus the degree of rehybridization $\tau$ decreases. 
Therefore, the energy of the defected structure increases with increasing $\theta$, as compared to the corresponding pristine structure (Fig. S4 in Supporting Information). 
Alternatively, the $\theta$-dependence can also be explained qualitatively by comparing the rotating C--C bond lengths for nanotubes with and without the defect, and we find the former to be shorter (Table S3 in Supporting Information). The difference $\Delta b$ ($= b_{P}-b_{SW}$) is larger for larger $\theta$: $\Delta b(\theta=0)<\Delta b(\theta=\pi/3$) for zigzag configurations, and $\Delta b(\theta=\pi/6)<\Delta b(\theta=\pi/2$) for armchair configurations. Thus,  the defected structure with $\theta=0$ ($\theta=\pi/6$)  is lower in energy, due to comparatively higher rehybridization, than the corresponding $\theta=\pi/3$ ($\theta=\pi/2$) structure.

\begin{figure}[t]
\centering
\includegraphics[scale=0.07]{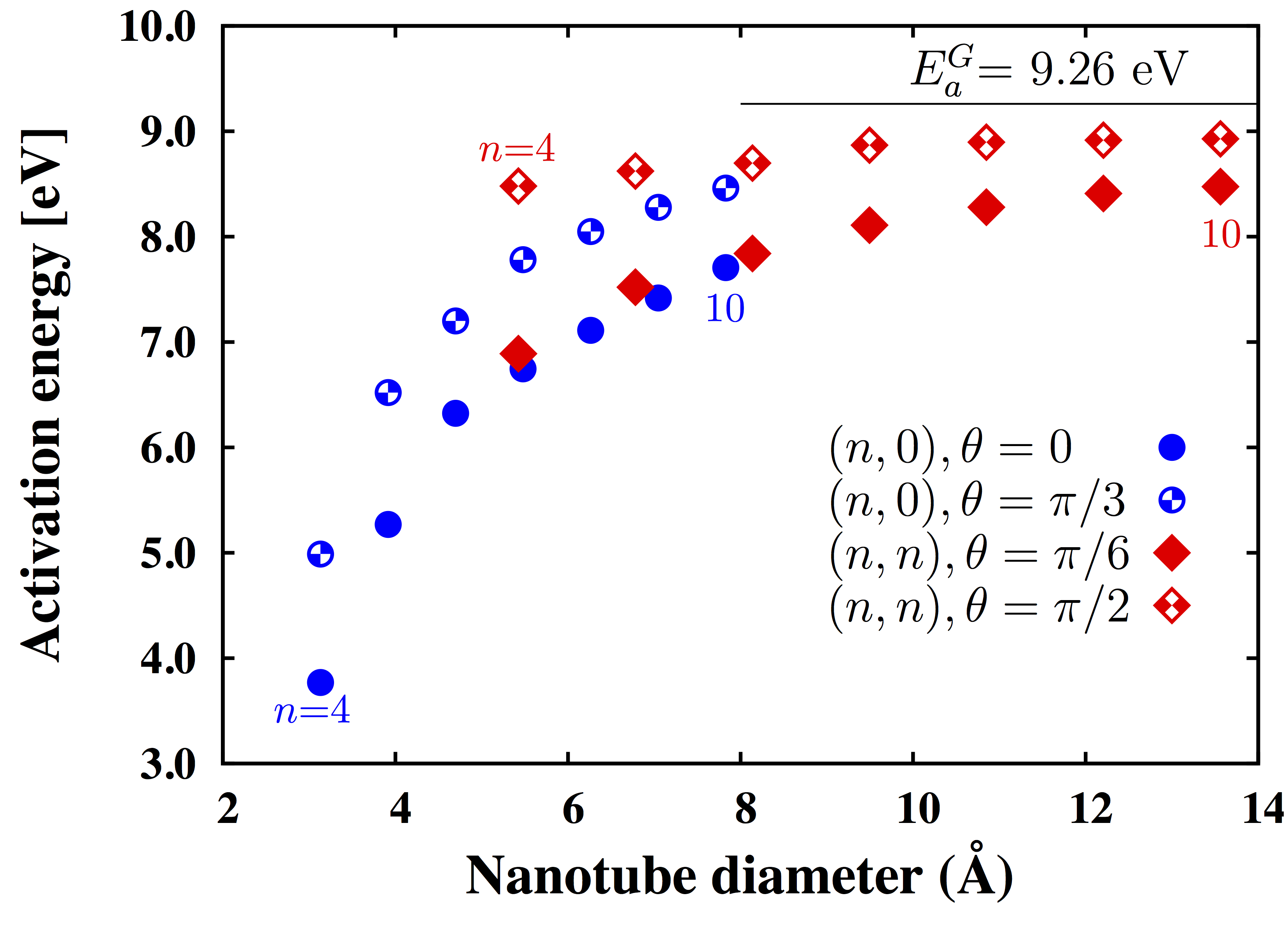}
\caption{Calculated activation barriers $E_a(d, \theta)$ show strong dependence on the chiral vectors (tube diameter $d$) and the defect orientation $\theta$. Such variation for zigzagg nanotubes is stronger compared to armchair counterparts. The SW activation barrier approaches that of the two-dimensional graphene (9.26 eV) with increasing tube diameter. The zero point energy $\Delta$ZPE correction lowers the activation barrier, which is found to be less than 0.25 eV for all cases.}
\label{fig:SWa}
\end{figure}

% We have E_f, why one would need E_a
{\bf Activation barrier, nanotube curvature, and defect orientation.} Thermodynamic quantities such as formation energy are insufficient to answer key questions of interest in CNT structural transformations. For example, how long does it take to form a metastable SW defect? Once formed, how long will such defects persist? This information related to formation kinetics under defined external conditions is important  to understand processes including mechanical deformation and CNT nanojunction or superstructure formation.  There are few estimates of this kinetic barrier to date, and none of which we are aware that considered potential orientation dependence on this barrier. Reported estimates have included incorrect descriptions of chemical bonding, and/or adopted methodologies to locate the first-order transition state that are now generally considered inadequate.~\cite{PhysRevB.57.R4277,dumitrica:2775}  Here, we locate the (first-order) transition state via DFT-CINEB methods described in Supporting Information, and subsequently calculate the corresponding activation barrier $E_a$.

The calculated $E_a (d, \theta)$ for varied tube diameter and inequivalent defect orientations are shown in Fig.~\ref{fig:SWa} for zigzag and armchair nanotubes. The overall qualitative trend of $E_a (d, \theta)$ is similar to that observed above for the formation energy: $E_a (d, \theta)$ increases with $d$, and shows a similar $\theta$ dependence (Fig.~\ref{fig:SWa}). For all cases considered, the calculated $E_a(d, \theta)$ converges to the graphene value $E_a^G$ (9.26 eV) with increasing $d$, and this convergence occurs at larger $d$ than for the formation energy [Fig.~\ref{fig:SWf}(e) and Fig.~\ref{fig:SWa}].  We calculated the activation energy for graphene $E_a^G$ with the identical theoretical approach, and also allowed the defect induced buckling perpendicular to the graphene plane. The present value for $E_a^G$ is in excellent agreement with previous calculations for graphene.~\cite{0957-4484-24-43-435707}  Similar to the trends observed for $E_f (d, \theta)$, for all tube diameters $E_a (\theta = \pi/3) > E_a (\theta = 0)$ for zigzag nanotubes and $E_a (\theta = \pi/2) > E_a (\theta = \pi/6)$ for armchair nanotubes. The complete $d$ and $\theta$ dependence of $E_a(d, \theta)$ can again be explained via curvature induced rehybridization. It is important to note that while the kinetic barrier of SW formation is very high (4--9 eV; Fig.~\ref{fig:SWa}), the reverse barrier [$E_a(d, \theta) - E_f(d, \theta)$] ranges between 4 and 5.5 eV for the nanotubes studied herein. This significant reverse activation barrier implies the (meta)stability of SW defects over a wide temperature range.

\begin{figure}[t]
\centering
\includegraphics[scale=0.07]{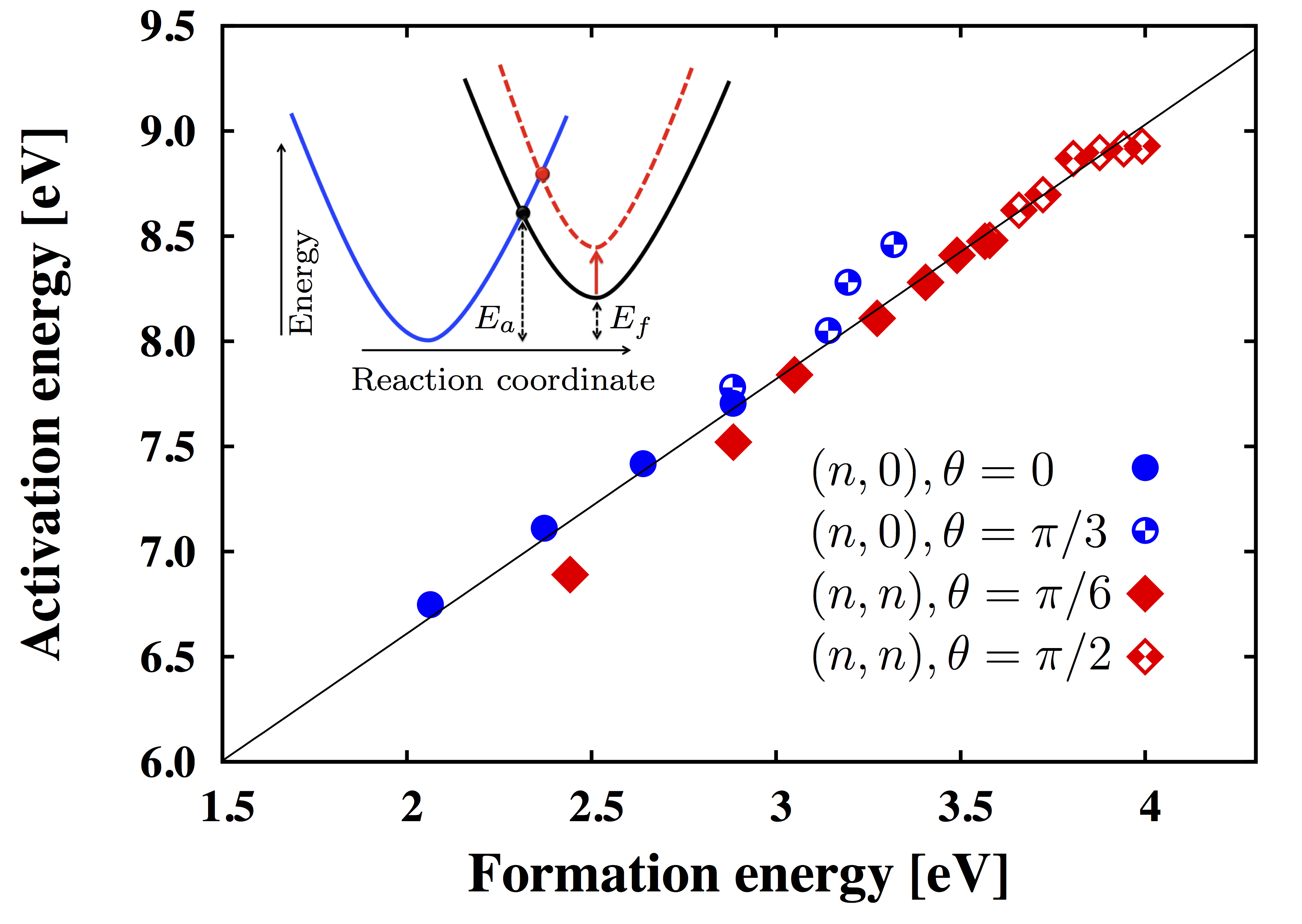}
\caption{Correlation between the thermodynamic formation energy and the kinetic barrier for $d > 0.5$ nm, which follows the linear Br{\o}nsted-Evans-Polanyi empirical rule.~\cite{Bronsted1928,Evans1938} The solid line is a linear fit, and the observed linear correlation can be explained by two overlapping parabolas representing the pristine and defected nanotube (inset).}
\label{fig:correlation}
\end{figure}

% Correlation between E_f and E_a
{\bf Correlation between ${\bm E_f}$ and ${\bm E_a}$.} This systematic study enabled investigation of possible and generalized correlations between the (thermodynamic) SW formation energy and the kinetic activation barrier. Indeed, we find an empirical Br{\o}nsted-Evans-Polanyi type linear relationship: $E_a = k_1 + k_2 E_f$, where $k$'s are empirical constants (Fig.~\ref{fig:correlation}).~\cite{Bronsted1928,Evans1938}  Although we have demonstrated above the capacity to locate transition states and calculate associated kinetic barriers in Fig.~\ref{fig:SWa}, that approach is computationally demanding and thus intractable for all possible combinations of CNTs and SW defect configurations. With this observed Br{\o}nsted-Evans-Polanyi correlation, we propose a reasonable estimate of activation energies that can be obtained for any nanotube via only knowledge of the formation energy (that is relatively easier to compute or measure). We find  $E_a = (4.19 \pm 0.15)$ eV  + $(1.05 \pm 0.04)E_f$ to be a good fit for the calculated values (Fig.~\ref{fig:correlation}). Such linear correlation between $E_a$ and $E_f$ can be understood qualitatively by two overlapping parabolas (inset of Fig.~\ref{fig:correlation}) representing the structures in the absence and presence of the defect.~\cite{AriehBook} In this model, if one or both of the parabolas shift in energy such that the energy difference between the minima (formation energy $E_f$) increases (decreases), the corresponding activation barrier concurrently increases (decreases).

% Manipulation of E_a  
{\bf Manipulation of activation barrier.} With this improved understanding of the relative thermodynamic and kinetic barriers of SW defects in CNTs, we next consider whether the considerable kinetic barrier for SW formation could be reduced significantly. Such a reduction that would promote SW defect formation is desirable in many novel electronic, mechanical and chemical applications, including the formation of CNT assemblies and superstructures.~\cite{PhysRevLett.92.075504, doi:10.1021/cr050569o,  Scientific.Report.2} It is known that applied uniaxial tension reduces the activation barrier.~\cite{PhysRevB.57.R4277,dumitrica:2775} Although that correlation explains the mechanical response of nanotubes, that approach to barrier reduction is not practically feasible for most applications.  Thus, here we assess other plausible ways to manipulate the activation barrier, and find that substitutional heteroatom doping (with elements B, N,  or S)  strongly modulates the activation barrier (Table~\ref{table1}). Regardless of the type of CNT (defined by chirality and diameter) and orientation of the defect, we find that the activation barrier is reduced substantially  ($\Delta E_a$ $\sim$ 1.3 - 4.6 eV) due to heteroatom doping at the active bond (Table~\ref{table1}). Doping with sulfur reduces the barrier most significantly, by 25--60\% depending on the tube type and defect orientation.  The reduction in activation barrier due to substitutional heteroatom doping can be qualitatively explained by bond weakening around the active site. This has been explained in detail earlier for fullerene.~\cite{PhysRevB.84.205404} Due to the weaker C--X bonds in X@CNT (X = B, N, S) compared to the C--C bonds in updoped CNTs, the SW rotation becomes easier for heteroatom doped CNTs. Such B, N and S-doped CNTs have been synthesized experimentally, and are proposed as metal-free electrocatalysts for oxygen reduction reactions.~\cite{Stephan09121994, Gong06022009, C3TA12647A} These dopants have also been found to facilitate the formation of novel three-dimensional CNT covalent networks,~\cite{doi:10.1021/nn700143q,ANIE:ANIE200705053,Scientific.Report.2} and our determination that SW defects are favored with such doping is consistent with such doping also favoring network formation.

\begin{table*}[t]
\caption{Heteroatom doping strongly influences the kinetic barrier for SW formation.  Calculated $E_a$ is reduced substantially due to B, N, and S doping, which is shown for semiconducting (10,0) and metallic (6,6) nanotubes with comparable diameters. The prefactor to the formation rate is calculated using harmonic transition state theory,~\cite{Vineyard1957121} $\nu = \Pi_i^{3N} \nu_i^{\rm P}/\Pi_i^{(3N-1)} \nu_i^{\rm TS}$, where $\nu_i^{\rm P} (\nu_i^{\rm TS})$ are the normal mode frequencies corresponding to pristine (transition state) structure.}
\begin{tabular}{rC{0.8cm}C{0.8cm}R{1.2cm}C{0.8cm}C{0.8cm}R{1.2cm}}
\hline
CNT & \multicolumn{2}{c}{Energy (eV)}& $\nu \times 10^{13}$ & \multicolumn{2}{c}{Energy (eV)} &$\nu \times 10^{13}$ \\
             &  $E_f$  & $E_a$ &  (Hz)                           &   $E_f$  & $E_a$ &  (Hz)\\
             \hline 
	    \hline 
	                   & \multicolumn{3}{c}{$\theta=0$} & \multicolumn{3}{c}{$\theta=\pi/3$} \\	
              (10,0)      & 2.88 & 7.71  &   229.3          &     3.32   & 8.44  &   93.8   \\
        B@(10,0)      & 2.01 & 5.77  &   195.3          &     3.37   & 6.37  &   17.3    \\               
        N@(10,0)      & 2.69 & 6.40  &    44.3           &      2.55   & 7.07 &   31.2    \\ 
        S@(10,0)      & 1.35 & 3.10  &     10.8          &     3.42   & 5.15   &   6.8    \\ 
             
             	           & \multicolumn{3}{c}{$\theta=\pi/6$} & \multicolumn{3}{c}{$\theta=\pi/2$} \\	
              (6,6)      & 3.15  & 7.84 &        19.1     &     3.72  & 8.70 &  42.9   \\
        B@(6,6)      & 2.52 & 5.63  &     31.9        &     3.96  & 7.24 &  6.0   \\
        N@(6,6)      & 2.57 & 5.56  &      16.3       &     2.87  & 7.03 &   21.6  \\
        S@(6,6)      & 1.92 & 3.25  &        7.1     &     3.79  & 6.59 &    4.6 \\

\hline              
\end{tabular}
\label{table1}
\end{table*}

The rate of SW defect formation can be estimated from the activation energy using a simple Arrhenius expression, $\Gamma = \nu \exp(-E_a/k_BT)$, where the prefactor $\nu$ is related to the vibrational frequency (Table~\ref{table1}),  $k_B$ is the Boltzmann constant, and $T$ is absolute temperature. Thus, with heteroatom doping the rate of SW activation becomes $\sim \exp(\Delta E_a/k_BT)$ times faster, as compared to the undoped case. For example, the rate of activation becomes six to 20 orders of magnitude faster for (10,0)-CNT due to heteroatom doping at temperatures relevant to fusion and chemical vapor deposition growth (1000 K). Thus, the reduction in activation barrier would promote CNT fusion via ion/electron irradiation,  as the fusion proceeds via a series of SW bond rotations.~\citep{doi:10.1021/nl015541g, PhysRevB.66.245403, PhysRevLett.92.075504,  NatMatKrasheninnikov} Moreover, the doping centers act as the SW nucleation center. This would be expected to facilitate covalent superstructure formation, which has been observed in recent experiments.~\cite{doi:10.1021/nn700143q,ANIE:ANIE200705053,Scientific.Report.2} The present calculation indeed supports these experimental observations, and further indicate that S-doping should be more effective in this regard because the reduction in activation barrier is much larger (Table~\ref{table1}). It is important to note here that due to the accelerated formation kinetics and increased thermodynamic concentration, the chemistry of SW defects should be accounted for accurately in such heteroatom-doped CNTs developed for catalytic applications.~\cite{Stephan09121994, Gong06022009, C3TA12647A}

\section{Conclusions}
In summary, we have studied the thermodynamic and kinetic properties of important topological Stone-Wales defects in single-wall carbon nanotubes, via density functional theory coupled with nudged elastic band identification of transition states. Calculated formation and activation barriers depend systematically on the tube chirality (and thus on tube diameter) and defect orientation. The microscopic origin of such dependence is attributable to curvature induced rehybridization. Generally, both the formation and activation energies increase with increasing (decreasing) tube diameter (curvature), and approach the respective values for two-dimensional graphene (Fig.~\ref{fig:SWf} and Fig.~\ref{fig:SWa}). The (kinetic) activation barrier is correlated with the (thermodynamic) formation energy, and follows the linear Br{\o}nsted-Evans-Polanyi relation (Fig.~\ref{fig:correlation}). Thus, the kinetic barrier for SW nucleation can now be  estimated from knowledge of the formation energy, the calculation of which is less demanding computationally. Further, we propose that the activation barrier can be manipulated substantially by heteroatom doping (Table~\ref{table1}) to increase the defect formation rate by up to 20 orders of magnitude at temperatures relevant to CNT fusion and chemical vapor deposition-based superstructure growth.  This computational finding explains the recent experimental observations that heteroatom doping favors CNT nanojunction and superstructure formation.~\cite{doi:10.1021/nn700143q,ANIE:ANIE200705053,Scientific.Report.2} Further, we propose that sulfur is a more effective dopant than nitrogen and boron for applications such as CNT fusion and superstructure formation that proceed via Stone-Wales bond rotation. The present findings can guide future experiments that seek to promote covalent CNT assembly.

%\begin{suppinfo}
Convergence of defect formation and activation energy on the tube length, and curvature dependent rehybridization and its effect on the formation and activation energy have been described and analyzed in the Supporting Information. Representative structures for heteroatom doped CNTs are also shown.
%\end{suppinfo}

%\begin{acknowledgement}
MK acknowledges helpful discussion with  A. Warshel, and a grant from the Department of Science and Technology, India under the Ramanujan Fellowship. Computational resources included the supercomputing facility at the Centre for Development of Advanced Computing, Pune, and at the Inter University Accelerator Centre, Delhi. The authors acknowledge use of computational resources supported by the National Research Foundation of Singapore through the BioSystems and Micromechanics Interdisciplinary Research Group of the Singapore-MIT Alliance for Research and Technology.
%\end{acknowledgement}

%\bibliography{cnt}

\providecommand{\latin}[1]{#1}
\providecommand*\mcitethebibliography{\thebibliography}
\csname @ifundefined\endcsname{endmcitethebibliography}
  {\let\endmcitethebibliography\endthebibliography}{}

\newpage

\section{Supporting Information}

\subsection{Defect orientation and tube-length}
The SW defects on the CNT surface can be generated at any of the three inequivalent sets of C--C bonds. Thus, for a particular CNT, there are three possible orientations of SW defect: $\theta =$ $\pi/3 - \chi$, $\chi$, and $\pi/3 + \chi$, where $\chi$ is the chiral angle. Out of these three sets, two of them are equivalent for both zigzag ($\chi = 0$) and armchair ($\chi=\pi/6$) nanotubes. The two inequivalent orientations are shown in the manuscript [Fig.~1 (a)-(b) for zigzag, and Fig.~1 (c)-(d) for armchair nanotubes].  Before any discussion of  formation energy $E_f$, and the kinetic barrier $E_a$ for SW defect, one should investigate their dependance on the tube length in the simulation box, which dictates the defect-defect interaction mediated via long-range strain field created by the defect itself. We calculated these quantities with varied tube length within the simulation cell, with $(6,m)$ CNTs as test cases (Table~\ref{table1}). We find that both $E_f$ and $E_a$ strongly depend on the tube length, and these calculated properties converge at lengths of 25.65 and 24.68 \AA ~for (6,0) and (6,6) nanotubes, respectively. Thus, throughout the present calculations, we consider these lengths for zigzag and armchair tubes, respectively, which minimizes defect-defect interaction. 

\begin{table}[!b]
\caption{Variation of formation energy $E_f$ and activation barrier $E_a$ with the length of the nanotube within the supercell. Due to the long-range nature of the strain field generated due to the topological defect, a large supercell along the tube axis is required to correctly predict the formation and activation energies.}
\begin{tabular}{cR{1cm}C{0.8cm}C{0.8cm}C{0.8cm}C{0.8cm}}
%\begin{tabular}{cC{1cm}C{1cm}C{1cm}C{1cm}C{1cm}}

\hline
($n,m$) & Length           & \multicolumn{2}{c}{Energy (eV)} & \multicolumn{2}{c}{Energy (eV)} \\
             &  (\AA)             &  $E_f$  & $E_a$                            &   $E_f$  & $E_a$  \\
             \hline 
	    \hline 
(6,0)       &          & \multicolumn{2}{c}{$\theta=0$} & \multicolumn{2}{c}{$\theta=\pi/3$} \\	
              &	        17.10     &	  2.01	 &	6.38	                     &	3.05		&  7.35 \\
              &         25.65     & 1.74		 &     6.32		             & 2.73		&  7.20  \\
              &         34.20     &	 1.72		 &     6.35		             & 2.72		&  7.14   \\

(6,6)       &          & \multicolumn{2}{c}{$\theta=\pi/6$} & \multicolumn{2}{c}{$\theta=\pi/2$} \\	
             
              &         9.87	      &	3.31		& 7.91                                &  4.40         &	8.95  \\
              &       14.80	      & 3.15		& 7.83	                      &  4.00	        &	8.91  \\
              &       19.74	      & 3.15		& 7.89                            &  3.72         &	8.70	  \\
              &       24.68	      & 3.10		& 7.82 	                     &  3.65	        &	8.75 \\       
\hline              
\end{tabular}
\label{table1}
\end{table}

\subsection{Phonon calculation}

\begin{figure}[!t]
\centering
\includegraphics[scale=0.015]{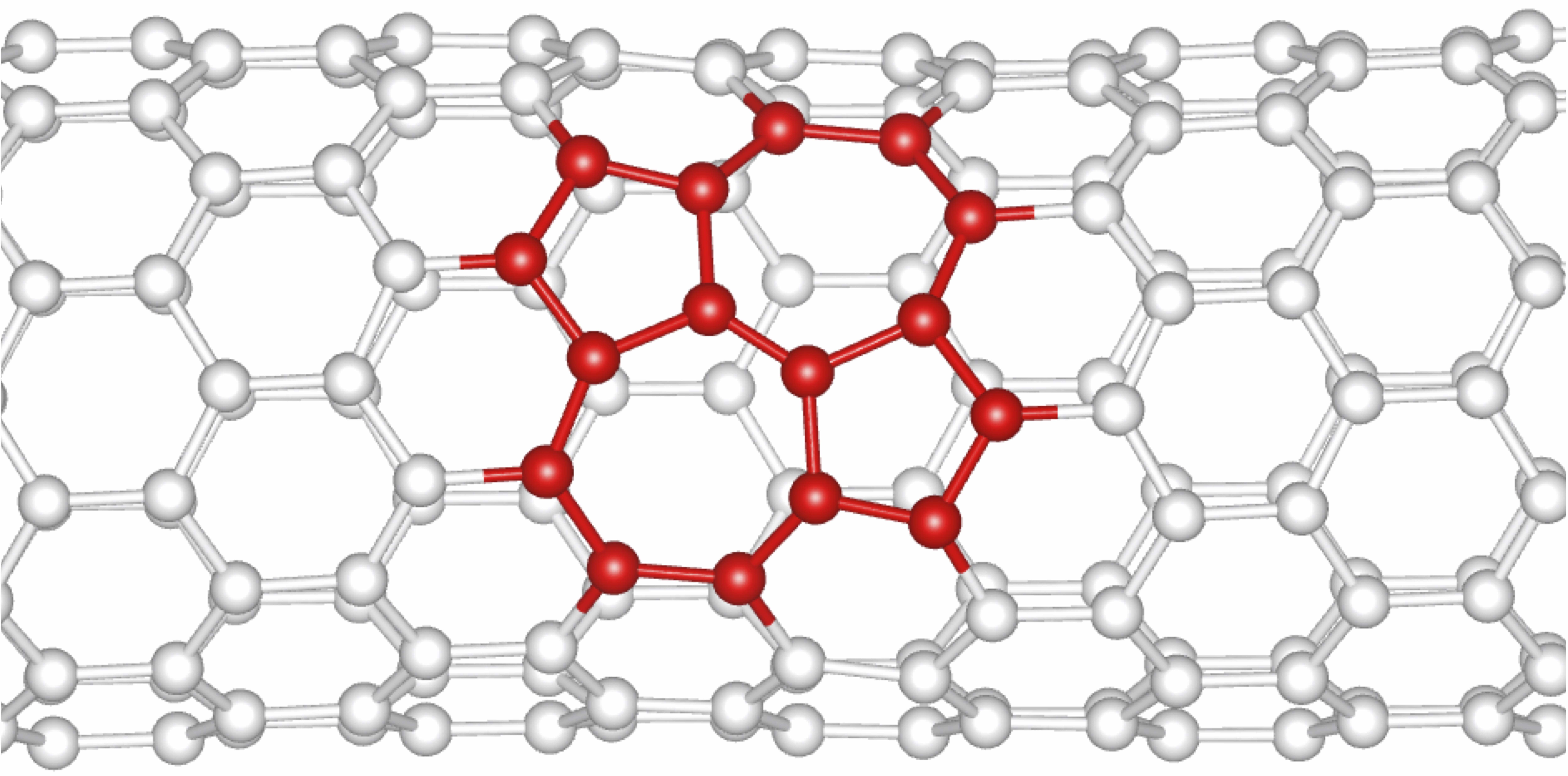}
\caption{Phonons are calculated by considering only the atoms around the rotating bond, which are affected by the C--C bond rotation. These atoms are highlighted in red color.}
\label{fig:phonon}
\end{figure}

Phonons are calculated using the finite difference method, in which we considered 16 atoms around the rotating bond that are mostly affected by the C--C bond rotation as highlighted in Fig.~\ref{fig:phonon}. All the transition states have been confirmed via one-and-only-one imaginary frequency, which indicate that these are indeed first-order transition states. These phonons are used to calculate the zero point energy correction to the formation energy and kinetic activation barrier.  The prefactor to the reaction rate within the harmonic transition state theory has been also calculated using these phonons:  $\nu = \Pi_i^{3N} \nu_i^{\rm P}/\Pi_i^{(3N-1)} \nu_i^{\rm TS}$, where $\nu_i^{\rm P} (\nu_i^{\rm TS})$ are the normal mode frequencies corresponding to pristine (transition state) structure.

\begin{figure}[!b]
\centering
\includegraphics[scale=0.12]{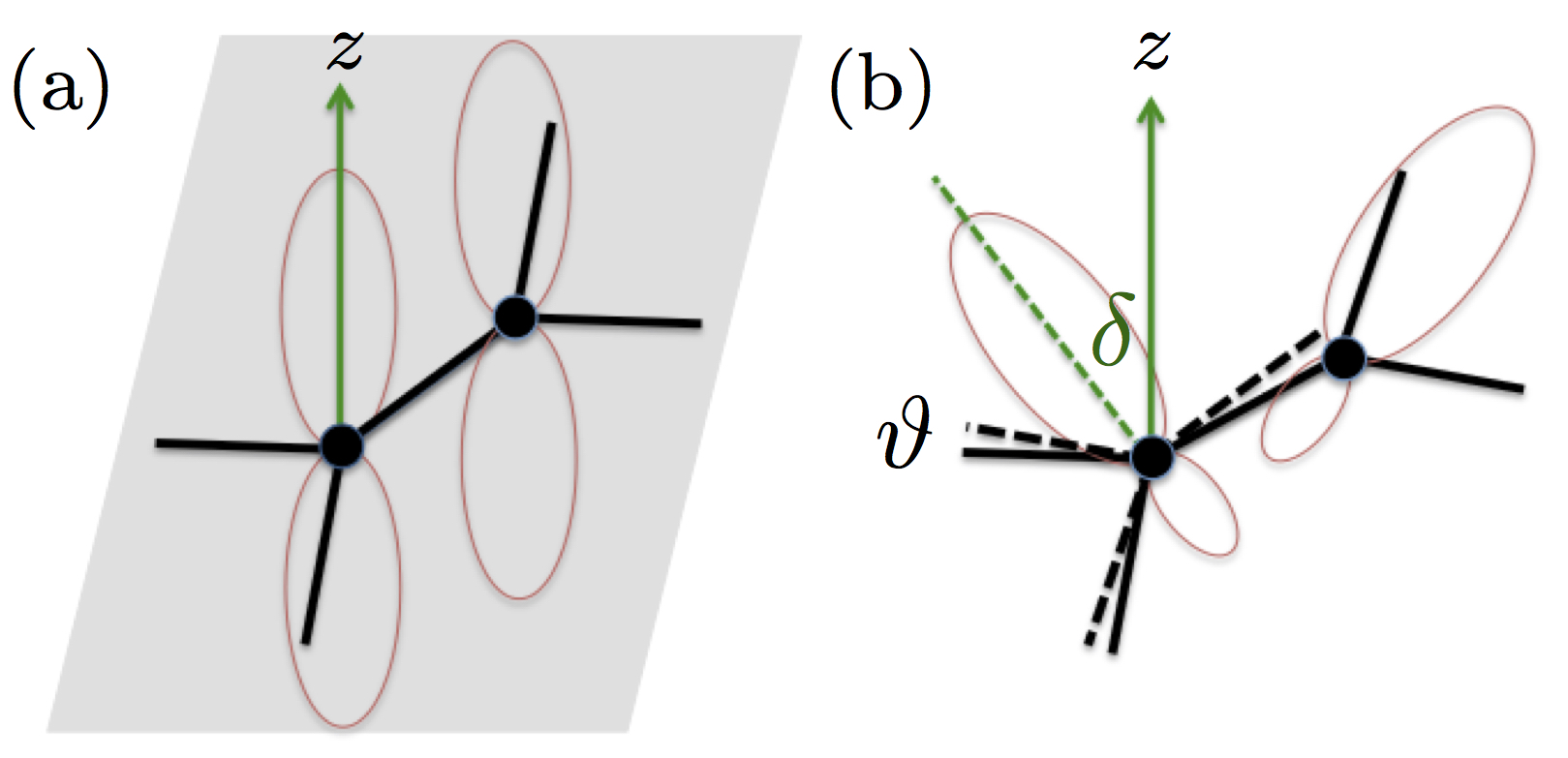}
\caption{(Color online) Schematic orbitals in graphene and nanotubes. (a) The $\pi$ orbital in planar graphene is orthogonal to the $\sigma$ bonds. (b) The $\pi$ orbital is no longer perpendicular to the $\sigma$ bonds on a curved CNT surface, as the $\sigma$ bonds are tilted down by an angle $\vartheta$ relative to the tangential direction of the tube, and the $\pi$ orbital bending by an angle $\delta$, respect to the normal drawn on tube surface. }
\label{fig:cartoon} 
\end{figure}

\subsection{Curvature and rehybridzation}
The carbon network on graphene is planar, and thus forms $sp^2$ hybridization with orthogonal $\sigma$ and $\pi$ orbitals [Fig.~\ref{fig:cartoon}(a)]. In contrast, the carbon atoms on the CNT surface lie on a curved surface, and thus the $\sigma$ bonds are pyramidalized, and the $\pi$ orbitals bend [Fig.~\ref{fig:cartoon}(b)]. Therefore, unlike in planar graphene, the $\sigma$ and $\pi$ orbitals are no longer perpendicular to each other. As a consequence the parts of the of the $\pi$ orbitals outside and inside rearrange due to Coulomb repulsion, and the outer contribution is much larger than the inner one [Fig.~\ref{fig:cartoon}(b)]. These lead to mixing of $\sigma$ and $\pi$ orbitals, which is known as rehybridization, and crucially depends on the nanotube diameter and chirality. The rehybridization leads to bonding which is in between $sp^2$ and $sp^3$, and can be recognized as $sp^{2+\tau}$ hybridization (where $\tau$ lies within 0 and 1, depending on the tube diameter and chirality).~\cite{Haddon17091993, Dumitrica2002182}

\begin{table}[!b]
\caption{Degree of rehybridization $\tau$ with tube diameter for zigzag and armchair nanotubes. Calculated $\tau$ decreases rapidly with increasing tube diameter $d$.}
\begin{tabular}{R{0.8cm}C{0.8cm}C{0.8cm}R{1.6cm}R{0.8cm}C{0.8cm}}
%\begin{tabular}{R{1cm}C{1cm}C{1.2cm}R{1cm}R{1cm}C{1.4cm}}

\hline\\
CNT & $d$ (\AA)   & $\tau$ & CNT & $d$ (\AA) & $\tau$ \\
\hline 
\hline
   (4,0) &       3.14 &      0.151 &    (4,4) &       5.44 &      0.051 \\
   (5,0) &       3.93 &      0.097 &    (5,5) &       6.80 &      0.033 \\
   (6,0) &       4.71 &      0.068 &    (6,6) &       8.16 &      0.023  \\
   (7,0) &       5.50 &      0.050 &    (7,7) &       9.53 &      0.017 \\
   (8,0) &       6.29 &      0.038 &    (8,8) &      10.89 &      0.013  \\
   (9,0) &       7.07 &      0.030 &    (9,9) &      12.25 &      0.010  \\
  (10,0) &       7.86 &      0.025 &   (10,10) &      13.61 &      0.008  \\
  \hline
\end{tabular}
\label{table:tau}
\end{table}

These facts can be mathematically accounted for within $\pi$ orbital axis vector construction, where it is assumed that the wave function is still separable in terms of $\sigma$ and $\pi$ orbitals. Assuming the $\sigma$ bonds are tilted down by an angle $\vartheta$ (pyramidalization angle) relative to the tangential direction of the tube. This introduces mixing of $p_z$ orbital with the $\sigma$ network. Under the orthogonality condition, the $\pi$ states on the curved nanotube surface can be written as,~\cite{Haddon17091993, Dumitrica2002182}
\begin{equation}
| h_{\pi}\rangle = \frac{1}{\sqrt{1+\lambda^2}}(|s\rangle + \lambda |p_z\rangle), 
\end{equation}
where $\lambda$ depends only on the pyramidalization angle $\vartheta$ as $\lambda = (1-3\sin^2\vartheta)/2\sin^2\vartheta$. Rehybridizied states have new wave functions, where $\pi$ orbital consists both $\sigma$ and $s$ orbitals. One can estimate the degree of rehybridization $\tau$ depending on the tube diameter $d$, and chirality $(n,m)$. Let $\delta$ be the bending angle of $\pi$ orbital relative to the normal drawn on the tube surface, and presuming that the angles between the $\sigma$ bonds and the $\pi$ orbitals are equal due to symmetry, one can show that $\delta$ depends on tube diameter, and chirality.~\cite{PhysRevB.64.113402,rehybridization} For a zigzag nanotube, 
\begin{equation}
\tan\delta = \frac{\sin^2\frac{\pi}{2n}}{\frac{\sqrt{3}\pi}{6n}+\sqrt{\frac{\pi^2}{12n^2}+\sin^2\frac{\pi}{2n}}}, 
\end{equation}
and for an armchair nanotube,
\begin{equation} 
\tan\delta = \frac{\tan\frac{\pi}{3n}\left( 2\sqrt{\frac{\pi^2}{12n^2}+\sin^2\frac{\pi}{6n}} - \tan\frac{\pi}{6n}  \right)}{2\sqrt{\frac{\pi^2}{12n^2}+\sin^2\frac{\pi}{6n}} + \tan\frac{\pi}{3n}}
\end{equation}

Finally, one can derive an analytical expression for the degree of rehybridization $\tau$ in $sp^{2+\tau}$ for both zigzag and armchair nanotubes,
\begin{equation}
\tau_{\rm zigzag} = \frac{4(1+3\sin^2\delta)}{3(1+2\sin^2\delta)} \frac{\sin^4 \frac{\pi}{2n}}{\frac{\pi^2}{12n^2}+\sin^2\frac{\pi}{2n}}, 
\end{equation}

\begin{equation}
\tau_{{\rm armchair}} = \frac{2(1+3\sin^2(\delta-\frac{\pi}{3n}))}{3(1+2\sin^2(\delta-\frac{\pi}{3n}))} \frac{\sin^2\frac{\pi}{3n}+2\sin^4\frac{\pi}{6n}}{\frac{\pi^2}{12n^2}+\sin^2\frac{\pi}{6n}}
\end{equation}

\begin{figure}[!t]
\centering
\includegraphics[scale=0.07]{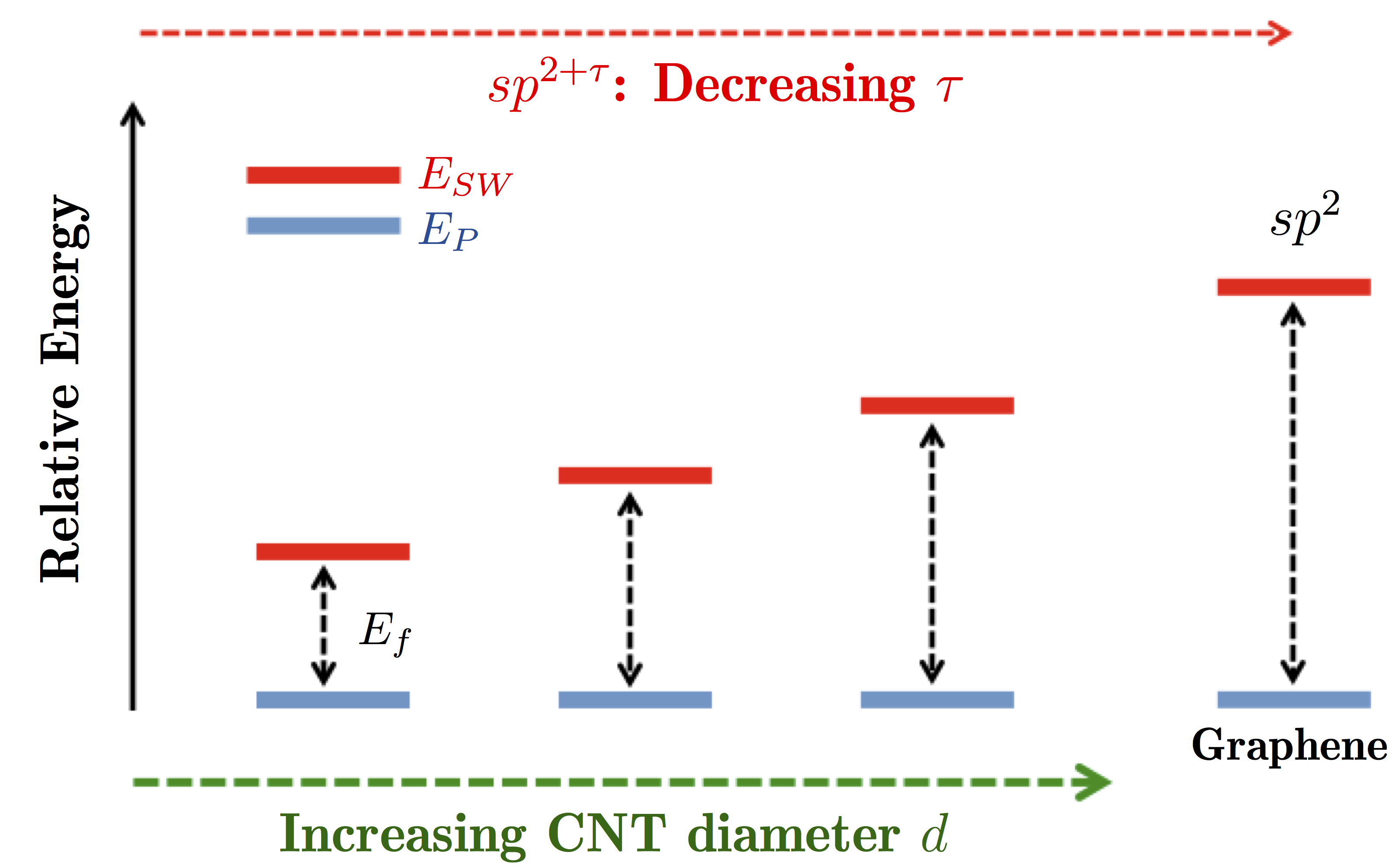}
\caption{The formation energy $E_f$ is calculated as the energy of the defected structure relative to the pristine tube. With the increase in tube diameter $d$, the curvature induced rehybridization decreases, and thus the true hybridization becomes more and more $sp^2$ like, and converge to pure $sp^2$ for flat graphene.  Thus, with increasing $d$ the defected structure lie higher in energy compared to the corresponding pristine tube, and therefore increasing the formation energy. Similarly, the activation energy also increases with increasing $d$.}
\label{fig:formation}
\end{figure}

\begin{figure}[!t]
\centering
\includegraphics[scale=0.07]{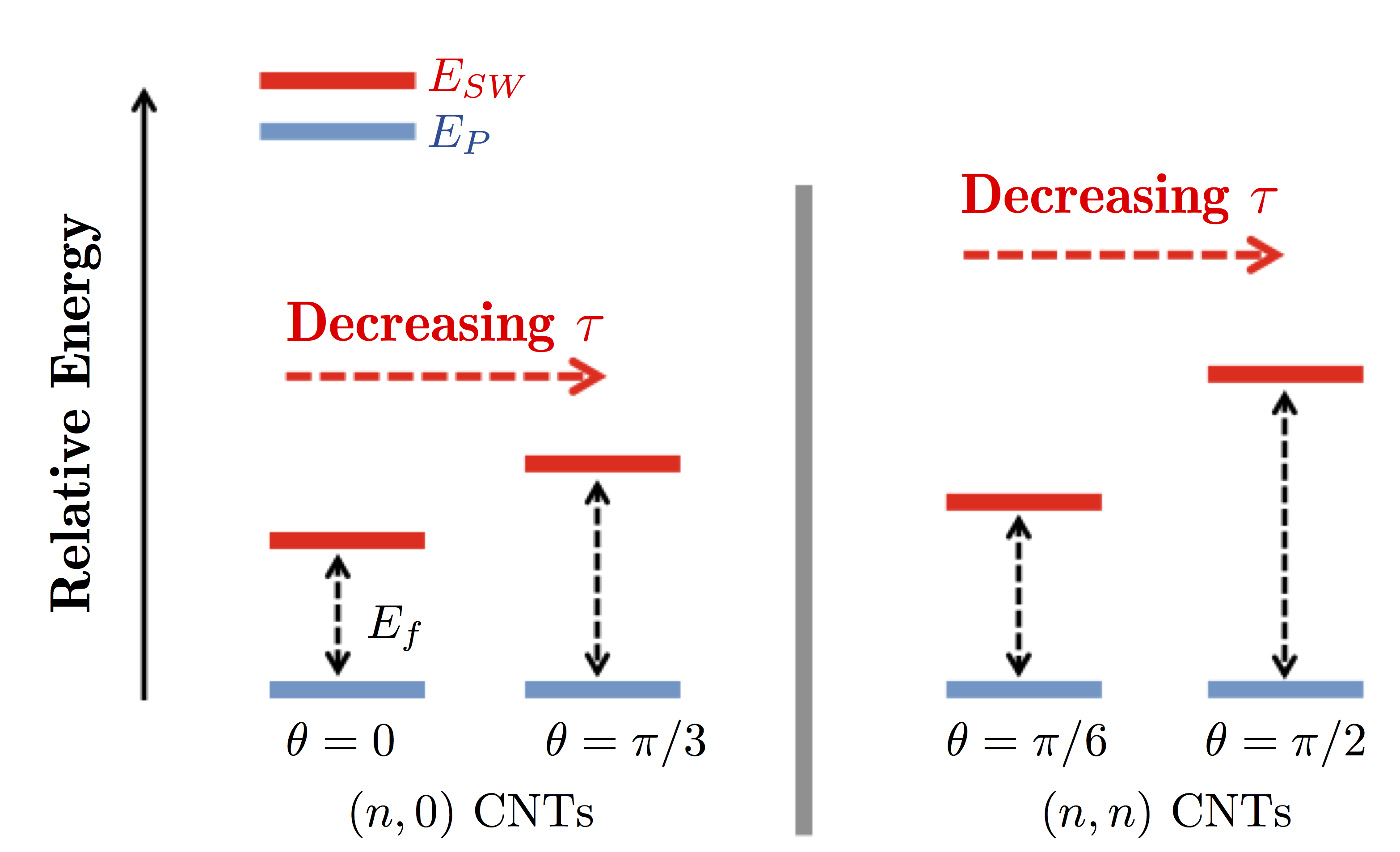}
\caption{For a particular CNT with diameter $d$, the rehybridization $\tau$ decreases with increasing $\theta$, the angle relative to the tube axis that is created by the rotating C--C bond in the pristine structure. For the defected structure, the local curvature of the rotated C--C bond increases with increasing $\theta$. Thus, $\tau$ decreases, which in turn increases $E_f$ and the corresponding $E_a$.}
\label{fig:orientation}
\end{figure}

{\tiny {
\begin{table*}[!t]
\caption{Carbon-carbon bond lengths before ($b_P$) and after rotation ($b_{SW}$) for zigzag and armchair nanotubes studied here. Bond lengths for different defect orientations are shown. All the values are in \AA. Difference in bond lengths between the pristine and defected structures ($\Delta b = b_p - b_{SW}$) are also shown.}
\begin{tabular}{C{0.8cm}C{0.8cm}C{0.8cm}C{0.8cm}C{0.8cm}C{0.8cm}C{0.8cm}|C{0.8cm}C{0.8cm}C{0.8cm}C{0.8cm}C{0.8cm}C{0.8cm}C{0.8cm}}
%\begin{tabular}{C{1cm}C{1cm}C{1cm}C{1cm}C{1cm}C{1cm}C{1cm}|C{1cm}C{1cm}C{1cm}C{1cm}C{1cm}C{1cm}C{1cm}}

\hline
\multicolumn{7}{c}{Zigzag nanotube} & \multicolumn{7}{c}{Armchair nanotube}  \\
CNT & \multicolumn{3}{c}{$\theta=0$} & \multicolumn{3}{c}{$\theta=\pi/3$} & CNT & \multicolumn{3}{c}{$\theta=\pi/6$} & \multicolumn{3}{c}{$\theta=\pi/2$} \\
              & $b_P$ & $b_{SW}$ & $\Delta b$ & $b_P$ & $b_{SW}$ & $\Delta b$ & & $b_P$ & $b_{SW}$ & $\Delta b$ & $b_P$ & $b_{SW}$ & $\Delta b$ \\ 
 \hline
 \hline
 (4,0)   &  1.400  &  1.417  & ---         & 1.484   & 1.386    & 0.098    & (4,4)    & 1.431    & 1.397     &  0.034   & 1.434    & 1.318     & 0.110  \\ 
 (5,0)   &  1.411  &  1.404   & 0.007   & 1.455   & 1.358    & 0.097    & (5,5)    & 1.429    & 1.379     & 0.050    & 1.429    & 1.321     & 0.108 \\
 (6,0)   &  1.413  & 1.390    & 0.023  & 1.447   & 1.362     & 0.085    & (6,6)   &  1.428    & 1.373     & 0.055    & 1.428    & 1.324     & 0.104 \\
 (7,0)   &  1.421  & 1.390    & 0.031  & 1.437   & 1.355    &  0.082   & (7,7)   &    1.427   & 1.366    &  0.061    & 1.427   & 1.325     & 0.102 \\
 (8,0)   &  1.419  & 1.386   & 0.033   &  1.436  & 1.349   &  0.087   & (8,8)  & 1.426       & 1.363     & 0.063     & 1.427   & 1.327    & 0.100 \\
 (9,0)   &  1.421  & 1.382   & 0.039   &  1.433  & 1.351 & 0.082     & (9,9)   & 1.426      & 1.360    & 0.066      & 1.426   & 1.328    & 0.098 \\
 (10,0) & 1.423   & 1.378   & 0.045  & 1.430   & 1.346  & 0.084    & (10,10) & 1.426    & 1.358    & 0.068      & 1.426   & 1.328    & 0.098 \\
 \hline
 \hline
\end{tabular}
\label{table:bond}
\end{table*}
}}

\begin{figure}[!b]
\centering
\includegraphics[scale=0.04]{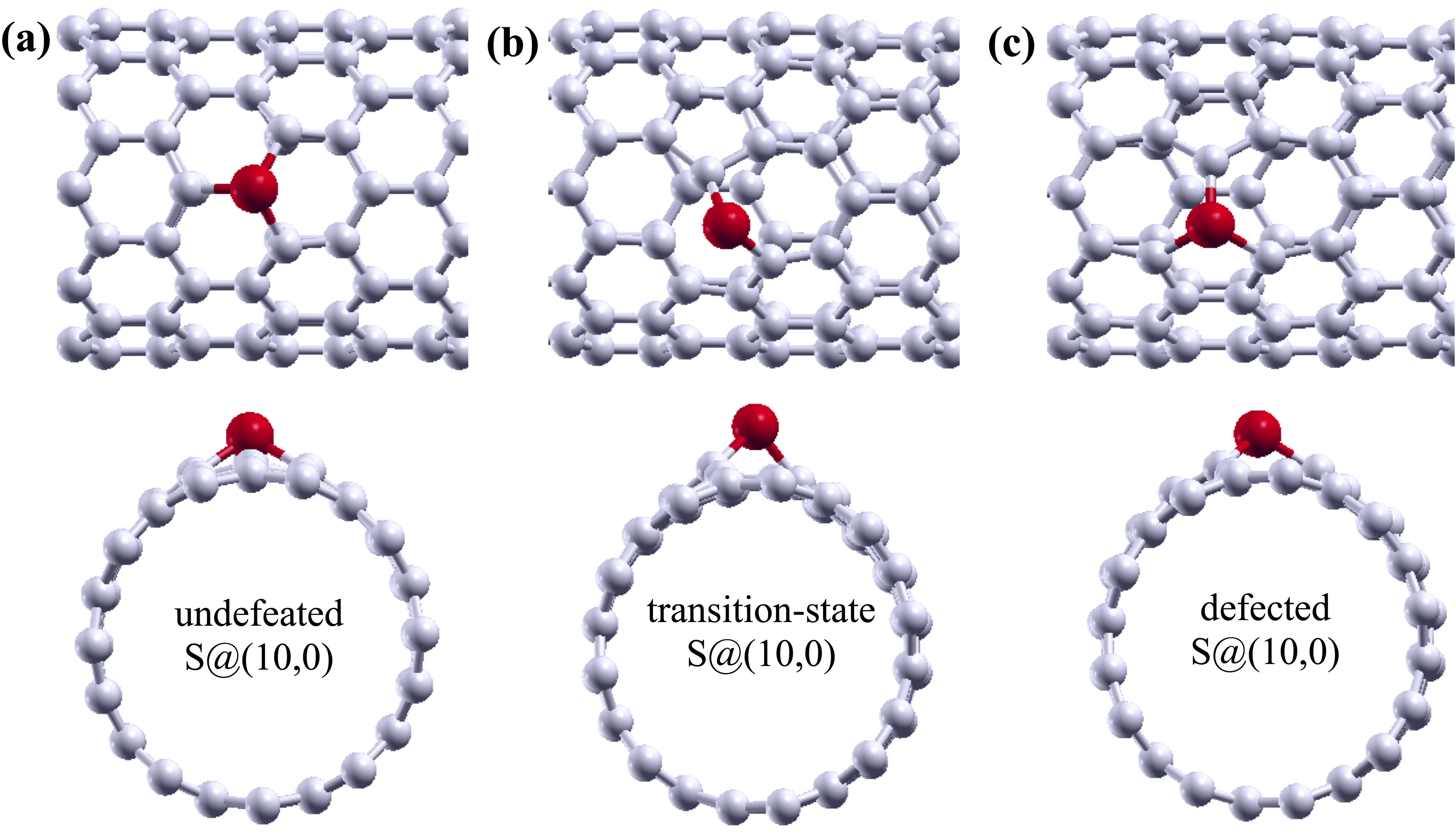}
\caption{Representative structures for heteroatom doped CNT. Here we show (a) undefected, (b) transition-state, and (c) SW defected geometries for S@(10,0) nanotube. The red ball is the substitutional S atom.}
\label{fig:doping}
\end{figure}

Calculated degree of rehybridization $\tau$ is shown in the Table~\ref{table:tau} for both zigzag and armchair nanotubes, and is clear that with the increase in tube diameter $d$ (i.e., with decreasing curvature) $\tau$ decreases monotonically, and will be zero for planar graphene.  Thus, with increasing diameter the true hybridization becomes more and more $sp^2$-like, as it has been discussed in the manuscript. This explains the $d$ dependance of both $E_f(d,\theta)$ and $E_a(d,\theta)$, which are calculated as the energy of the defected structure or the transition-state, respectively, relative to the pristine tube. Figure~\ref{fig:formation} explains the $d$ dependance of  $E_f(d,\theta)$. With increasing $d$, the rehybridization $\tau$ of the rotated C-C bond in the defected structure decreases, and thus the defected structure is pushed toward higher energy relative to the pristine structure. Thus, increasing the energy difference between the structure with and without the defect explains the observed increasing $E_f(d,\theta)$ with increasing $d$. Similarly, the $d$ dependance of $E_a(d,\theta)$ can be explained by considering the first-order transition state.

The above explanation does not account for the local curvature of the rotating/rotated C--C bond, as it was assumed that all three $\sigma$ bonds are equal and are all are tilted down equally. Ot was also assumed that the $\pi$ orbitals form equal angles with the $\sigma$ bonds. However, this is not the case, specially for the tubes with smaller diameter. This fact is evident from the bond length analysis shown in Table~\ref{table:bond}. Depending on the orientation of the rotating C--C bond, bond length differs reflecting the effect of curvature induced rehybridization. However, this difference decreases with increasing tube diameter. The theta dependence can be explained by considering these effect of these features on the local curvature of the rotated C--C bond. With increasing $\theta$ the local curvature of rotated C--C bond decreases, and consequently the rehybridization $\tau$ decreases (shown in Fig.~\ref{fig:orientation}). Therefore, both the formation energy and activation barrier increase with increasing $\theta$ for both zigzag and armchair nanotubes.    

% \theta dependence is more prominent in tubes with smaller diameter. 

\subsection{Substitutional doping}
Substitutional heteroatom (such as B, N, and S) doped CNTs have been experimentally synthesized. The topological SW defect activation is much easier in these CNTs.  Representative undefected, transition-state and SW defected structures are shown for S@(10,0) nanotube in Fig.~\ref{fig:doping}. In this case one could calculate two different formation energies. The energy requirement for heteroatom substitution, i.e., the formation energy of the substitutional defect.~\cite{reviewer} However, we are not interested in this formation energy, as our goal is to calculate the SW defect formation energy once we already have doped CNT. The SW defected formation energy is calculated as, $E_f({\rm X@CNT}) = E_{{\rm SW}}({\rm X@CNT}) - E_{\rm P}(\rm X@CNT)$, where X=B, N, or S. The reduction in SW activation barrier for X@CNTs can be understood by bond weakening around the active site. The corresponding X--C bonds are much weaker than the C--C bonds. For example, the S--C bond strength (2.73 eV) is much weaker than C--C bonds (5.18 eV) for S@(10,0) nanotube. Thus, the bond rotations become much easier in X@CNTs.  

%\bibliography{cnt}

\providecommand{\latin}[1]{#1}
\providecommand*\mcitethebibliography{\thebibliography}
\csname @ifundefined\endcsname{endmcitethebibliography}
  {\let\endmcitethebibliography\endthebibliography}{}

\end{document}